\documentclass[twocolumn,showpacs,preprintnumbers,amsmath,amssymb]{revtex4}
\usepackage{graphicx}% Include figure files
\usepackage{dcolumn}% Align table columns on decimal point
\usepackage{bm}% bold math

\begin{document}
\def\al{&\!\!\!\!}
\def\x{{\bf x}}
\def\f{\frac}
\def\y{\frac{1}{2}}
\preprint{APS/123-QED}

\title{Dark companion of baryonic
matter. III}
% Force line breaks with \\

\author{Y. Sobouti}
 \altaffiliation[ ]{Institute for Advanced Studies in Basic Sciences - Zanjan, Iran}
% Lines break automatically or can be forced with \\
 \email{sobouti@iasbs.ac.ir}
\affiliation{%
Institute for Advanced Studies in Basic Sciences-Zanjan, Iran}
%\affiliation{

\date{\today}% It is always \today, today,

\begin{abstract}
Wherever one talks of dark matter, one does so where there is an
observable matter and an associated unsolved dynamical issue to be
settled. We promote this observation to the status of an axiom and
conjecture that there is a dark companion to every baryonic
matter, subject to certain rules as regards its size,
distribution. To pursue the proposition in a systematic way we
resort to the rotation curves of spiral galaxies. They have non
classical features. First, we design a spacetime metric around the
galaxy to accommodate these  features. Next we calculate the
density and pressure of a hypothetical dark matter that could
generate such a spacetime. In the weak field regime and for a
spherical distribution of mass $M$, we are able to assign a dark
perfect gas companion, whose density is almost proportional to
$M^{1/2}$ and fades away almost as $r^{-2}$. However, in view of
this orderly relation between the observable mass and its dark
companion, one may choose to interpret the whole scenario as an
alternative theory of gravitation.
\end{abstract}

\pacs{Valid PACS appear here}% PACS, the Physics and Astronomy
\maketitle

\section{\label{sec:level1}Introduction}
That the baryonic content of galaxies, cluster of galaxies, or for
that matter the universe at large, does not provide sufficient
gravitation to explain the observed dynamics of the systems is an
established fact. To resolve the dilemma, dark matter (energy)
scenarios and/or alternative theories of gravitation have been
speculated and debated. The fact, however, remains that the
proponent of dark matter (energy) have always looked for it in
baryonic environments. No one  has, so far, reported a case where
there is no ordinary matter, but there is a dynamical puzzle  to
be resolved. In view of this negative observation, it is not
unreasonable to conjecture that:

`Any baryonic matter has an ever attendant dark companion, and
there are rule to this companionship as regards the size and the
distribution of the matter and its twin companion.'

On the other hand such a point of view, that denies the
independent existence of the dark matter, is equivalent to the
assumption that the known theories of gravitation, based on
baryonic matter alone, do not tell the whole story and there is
room for amendments.  This conclusion in turn reduces the
distinction between the dark matter scenarios and alternative
theories to the level of semantics: As long as the dark matter
betrays no interaction with the baryonic matter other than the
gravitational one, one has the option either to assume a dark
component to every baryonic matter subject to certain rules, and
account for its gravitational field in the conventional way; or
simply adhere to the baryonic matter but come up with an
alternative law of gravitation.

This paper, like its two precursors \cite{sob1}, \cite{sob2}, is
an inverse approach to understand the idiosyncracies of the
rotation curves of spiral galaxies. Based on observations we first
design a spacetime metric that is capable of supporting the
non-classical features of the rotation curves. This step amounts
to actually giving the gravitational potential at the outer
reaches of a galaxy. Next we attribute the deviations from the
conventional baryon induced gravitation to a dark companion to the
galaxy, and give a rule for the size and distribution of its
density and pressure.

\section{Observed facts and implication}
There are three main characteristics to the rotation curves of
spirals
\begin{itemize}
\item They decline, if at all, much less steeply than the Keplerian curves do,
see, e. g. \cite{sho73} -
%, \cite{bos83}, \cite{beg89}, \cite{bbs91},
%\cite{sv98}, \cite{smcg02}, \cite{ps95}, and
\cite{pss96}.
   \item Beyond the visible disks of the galaxies, orbital speeds
are, more often than not, proportional to the fourth root of the
mass
    of the galaxy, the Tully-Fisher relation \cite{tf77}.
\item Deviation from  the classical concepts, in this case
        the gravitation, show up in large scale systems and at large distances,
        or in the description of Milgrom at small gravitational accelerations
\cite{mil83abc}.
\end{itemize}
These observed facts are our starting points. The galaxy, though a
flattened system, is approximated by a spherically symmetric
distribution of baryonic matter. Accordingly the spacetime
external to it will be static and spherically symmetric:
\begin{eqnarray}
ds^2=-B(r)dr^2+A(r)dr^2+r^2(d\theta^2+\sin^2\theta d
\varphi^2).\label{eq1}
\end{eqnarray}
We adopt a dark matter language and assume that the galaxy
processes a static dark perfect gas companion of density
$\rho_d(r)$, of pressure $p_d(r)<< \rho_d(r)$, and of 4-velocity,
$$U_t=-B^{1/2}, ~U_r=U_\theta=U_{\varphi}=0.$$ In the baryonic
vacuum, Einstein's field equations become
\begin{eqnarray}
R_{\mu\nu}-\f{1}{2}g_{\mu\nu}R=-\left[p_d~g_{\mu\nu}+(p_d+\rho_d)U_\mu
U_\nu\right],\label{eq2}
\end{eqnarray}
where we have let $8\pi G$ and $c^2$ equal to 1. To respect the
Bianchi identities, one must require the 4-divergence of the right
hand side of Eq. (\ref{eq2}) to varnish. This, in turn, leads to
the hydrostatic equilibrium of the dark fluid and to a
differential equation for $p_d$. From Eq. (\ref{eq2}) the two
combinations, $$R_{tt}/B+R_{rr}/A+2R_{\theta\theta}/r^2
~\textrm{and}~ R_{tt}/B+R_{rr}/A,$$ give
\begin{eqnarray}
\al\al \f{1}{r^2}\left[\f{d}{dr}\left(\f{r}{A}\right)-1\right]=-\rho_d,\label{eq3} \\
\al\al \f{1}{r
A}\left(\f{B'}{B}+\f{A'}{A}\right)=\rho_d+p_d,\label{eq4}
\end{eqnarray}
respectively. To the first order of smallness, we neglect $p_d$ in
comparison with $\rho_d$, eliminate $\rho_d$ between the two
equations, and find
\begin{eqnarray}
\f{B'}{B}=\f{1}{r}(A-1).\label{eq5}
\end{eqnarray}
We now assume $(A-1)$ is analytic and differentiable, and has the
following series expansion at large $r$'s:
\begin{eqnarray}
A(r)-1=\left(\f{r_0}{r}\right)^\alpha
\sum_{n=0}\f{s_n}{r^n},\label{eq6}
\end{eqnarray}
where the indicial exponent, $\alpha$, is dimensionless, and $r_0$
is an arbitrary length scale, presumably of the order of the size
of the galaxy. The constant parameters, $s_n$, are of dimension
$(\textrm{length})^n$. Their size will be discussed shortly. It
should be pointed out that the expansion of Eq (\ref{eq6}) is for
regions external to the galaxy. In particular, the center $r=0$ is
not included in the indented domain and no question of singularity
will arise.

Next we substitute Eq. (\ref{eq6}) in Eq. (\ref{eq5}) and
integrate for $B(r)$. Depending on whether $\alpha$ is zero or
not, two different solutions emerge:
\begin{eqnarray}
B \al=\al~ \exp\left[-\left(\f{r_0}{r}\right)^\alpha\sum_{n=0}\f{s_n}{(n+\alpha) r^n}\right],~~~~\textrm{for}~ \alpha\neq 0,\label{eq7} \\
\al=\al
\left(\f{r}{r_0}\right)^{s_0}\exp\left[-\sum_{n=1}\f{s_n}{nr^n}\right],\hspace{1.2cm}\textrm{for}~
\alpha=0.\label{eq8}
\end{eqnarray}
Both gravitation- and speed-wise, galactic environments are non
relativistic. In the weak field regime, we retain the first two
terms in the series expansion of the exponential terms,
approximate the gravitational potential by
$\phi(r)=\f{1}{2}(B-1)$, calculate the circular speed of a test
object orbiting the galaxy from $v^2 = rd\phi/dr$, and find
\begin{eqnarray}
 v^2\al =\al
 %\f{1}{2}\left(\f{r_0}{r}\right)^\alpha \sum_{n=0}\f{s_n}{r^n} =
 \f{1}{2}\left(\f{r_0}{r}\right)^\alpha\left[s_0+\f{s_1}{r}
 +\cdots\right], ~~~~\alpha\neq0,\label{eq9}\\
\al = \al
%\f{1}{2}\left(\f{r_0}{r}\right)^{s_0} \sum_{n=0}\f{s_n}{r^n} =
\f{1}{2}\left(\f{r}{r_0}\right)^{s_0}\left[s_0+\f{s_1}{r}
+\cdots\right], ~~~~\alpha=0.\label{eq10}
\end{eqnarray}
At large distances from the galaxy, the $s_0$ terms in Eqs.
(\ref{eq9}) and (\ref{eq10}) are the dominant ones. As one moves
closer, $s_1$ terms gain over $s_0$. Further inward, $s_2$ and
higher terms may take turn. It should, however, be noted the
formalism devised here is to deal with velocity  anomalies at the
outer  reaches of the galaxies, beyond their visible extensions.
No chance will arise for $s_2$ and higher order terms to play
roles.

Logarithmic slopes of the rotation curves, $\Delta=d\ln v^2/d\ln
r$, and their asymptotic behaviors are

\begin{eqnarray}
\textrm{a})~ ~~ \Delta\al = \al \left[-\alpha
s_0-(1+\alpha)\f{s_1}{r}\right]/\left[s_0+\f{s_1}{r}\right],~~\alpha\neq
0~~~~~\cr \al \rightarrow \al -\alpha~ ~~~~~~~~~\textrm{at
~large}~ r \cr
\al \rightarrow \al -(1+\alpha)~~~\textrm{at~ small}~ r. \label{eq11}\\
~\nonumber\\
 \textrm{b})~ ~~ \Delta\al = \al
\left[s_0^2-(1-s_0)\f{s_1}{r}\right]/\left[s_0+\f{s_1}{r}\right],~~~~~~\alpha=0~~~~~
\cr \al \rightarrow \al s_0 ~~~~~~~~~~~~~\textrm{at ~large} ~r \cr
\al \rightarrow \al - (1-s_0)~~\textrm{ at~ small}~ r.
\label{eq12}
\end{eqnarray}

 At far distances, one has the falling slope $-\alpha$ in case (a),
and the rising slope $s_0$ in case (b). In either case one,
however, knows that the observed asymptotic slopes are much less
steep than the Keplerian slope, $-1$. Therefore, $\alpha$ in case
(a) and $s_0$ in case (b) should be much smaller than 1. At closer
distances the slopes are almost Keplerian, except for small
abberation by $+\alpha$ or $-s_0$. This, however, should not be
taken seriously. It is a consequence of the assumption of
spherical symmetry of the model and is not expected to be present
in actual flat galaxies.

\section{Determination of $\alpha$ and $s_n$'s}

$\alpha$):  A study of the asymptotic slopes of the rotation
curves of spirals can, in principle, give $\alpha$ of Eq.
(\ref{eq11}), if the slope is negative, or $s_0$ of Eq.
(\ref{eq12}) if it is positive. Here, however, we are content with
an order of magnitude of these parameters. In their list of 1100
rotation curves, primarily used to derive a universal rotation
curve, Persic et al \cite{pss96} find a subset of 27 reliable
curves extending out to $2 R_{\textrm{optical}}$ and 200
statistically significant ones farther out than
$R_{\textrm{optical}}$. They define a dark matter indicator, $
\delta=[v(2R_{\textrm{opt}})-v(R_{\textrm{opt}})]/v(R_{\textrm{opt}})
$, which is almost one half of the slopes of Eqs. (\ref{eq11}) and
(\ref{eq12}), in the interval $(1-2)R_\textrm{opt}$. Thus
\begin{eqnarray}
\Delta(R_\textrm{opt}) \approx 2 \delta=-2[0.05+0.16 \log (L/10^{10.4}L_\odot)],
\label{eq13}
\end{eqnarray}
where $L$ is the luminosity of the galaxy. They note, the
expression is valid in the magnitude range $-23.2<M_I<-18.5$.
Depending on $L$, negative $\Delta$'s are roughly, $\alpha$ of Eq.
(\ref{eq11}) and the positive ones are $s_0$ of Eq. (\ref{eq12}).
In either case they are small and fall in the range of $\pm 10$
percent. Actual \emph{`asymptotic'} slopes, however, should be
well below what Eq. (\ref{eq13}) indicates. For, one can hardly
convince oneself that rotation curves at distances of $(1.5-2
)R_{\textrm{opt}}$ have actually reached their asymptotic regime.
To summarize, we are inclined, after a qualitative examination of
a good number of rotation curves in \cite{ps95}, \cite{pss96},
\cite{smcg02}, \cite{tc09} and others, to infer from Eq.
(\ref{eq13}) the value `$\alpha\leq \textrm{few percents}$' for
the negative asymptotic slopes of Eq. (\ref{eq11}), with a fair
confidence. The case of asymptotically  positive slopes, if they
occur in nature at all,
is discussed below.\\

$s_0$): The Tully-Fisher relation, initially a power law
expression between the circular rotation speeds at the outer
reaches of spiral galaxies and their luminosities, can be
expressed as a power law relation between the asymptotic speeds
and galactic masses. Thus, $v_\infty\propto M^\beta$. A range of
values for the exponent, $\beta$, can be found in the literature
\cite{mc00}, \cite{gmetal04}, and \cite{tc09}. We adhere to the
commonly quoted value $\beta=1/4$.

The length scale $r_0$ in Eqs. (\ref{eq9}) and (\ref{eq10}) is
arbitrary. We choose it roughly the distance to which the rotation
curves are extended out to. Because of the smallness of $\alpha$
and $s_0$ the factors $(r_0/r)^\alpha$ and $(r/r_0)^{s_0}$
approximate to 1 and the asymptotic speed in both equations
becomes $v_\infty^2\approx s_0/2$. This, by Tully-Fisher relation,
gives
 \begin{eqnarray}
s_0=\lambda\left(\f{M}{M_\odot}\right)^{1/2} \label{eq14}
\end{eqnarray}
where $M$ is the baryonic mass of the galaxy, stars+gas, etc. To
determine the proportionality constant, $\lambda$, one turns to
observations.  From  a list of 31 galaxies in \cite{smcg02}, we
find $\lambda\approx 3\times 10^{-12}$ \cite{sob07}. A better
estimate is available through MOND. The weak acceleration limit of
MOND  \cite{mil83abc} is
$$v^2/r= \f{1}{2}\lambda(M/M_\odot)^{1/2}c^2/r=
(a_0GM/r^2)^{1/2},$$  where we have restored the factor $c^2$
which was suppressed so far, and $a_0=1.2\times 10^{-8} \rm{cm~
sec}^{-2}$ \cite{beg89} is the universal acceleration of MOND.
This yields
\begin{eqnarray}
\lambda = 2(GM_\odot a_0)^{1/2}/c^2 = 2.8 \times
10^{-12}.\label{eq15}
\end{eqnarray}

$s_1)$: The term $ s_1/2r$ in Eqs. (\ref{eq9}) and (\ref{eq10}),
operative at closer distances, is actually the classical newtonian
term. Thus, $s_1$ should be identified with the Schwarzschild
radius of the galactic mass:
\begin{eqnarray}
s_1 = 2GM/c^2. \label{eq16}
\end{eqnarray}
   For the remaining $s_2$ and higher terms we have no suggestion
   at present.  If they exist at all, our idealized
   spherically symmetric model does not sufficiently closely mimic
   the actual flattened
   galaxies to draw meaningful conclusions.

\section{ The dark companion}

From Eqs. (\ref {eq6}) and (\ref{eq3}) the density of the dark
fluid is
\begin{eqnarray}
\rho_d = \left(\f{r_0}{r}\right)^\alpha \f{1}{r^2}\left[
(1-\alpha) \lambda \left(\f{M}{M_\odot}\right)^{1/2} -
\alpha\f{s_1}{r}\right],\label{eq17}
\end{eqnarray}
where we have substituted for $s_0$ from Eq. (\ref{eq14}). The
expression is valid for $\alpha=0$ as well.  For all practical
purposes, the second term in the bracket can be neglected.  That
the dark density is almost proportional to the square root of the
mass of the galaxy is a direct consequence of the Tully-Fisher
relation. That it fades away exactly or approximately as $r^{-2}$,
depending on whether $\alpha = 0$ or not, is in accord with
$\Lambda \rm{CDM}$
 simulations of \cite{nfw97} and others.

 The dark matter inside a  radius $r$, $M_d(r)= 4\pi\int{\rho r^2
 dr}$, is
\begin{eqnarray}
M_d(r)= 4\pi\lambda\left(\f{M}{M_\odot}\right)^{1/2}
\left(\f{r_0}{r}\right)^\alpha r. \label{18}
\end{eqnarray}
It is said that the role of the dark matter is more prominent in
intrinsically fainter galaxies than in brighter ones, see e. g.
\cite{pss96}. This is inferred from  the fact that deviations of
the actual rotation curves from the baryonic-matter-based
Keplerian ones is larger in fainter galaxies than in the brighter
ones. This can be understood by considering the ratio $
M_d(r)/M\propto r^{(1-\alpha)} M^{(-1/2)}$. At a given $r$,
normalized to the optical radius of the galaxy, say, this ratio
decreases as the galactic mass or equivalently its luminosity
increases.

The pressure of the dark fluid is obtained by letting the 4-
divergence of the right hand side of  Eq. (\ref{eq2}) vanish. it
leads to the hydrostatic equilibrium of the dark matter:
\begin{eqnarray}
\f{p_d'}{p_d+\rho_d}\approx \f{p_d'}{\rho_d}=
-\f{1}{2r}(A-1).\label{eq19}
\end{eqnarray}
Integration is straightforward.  The first two terms in the series
are
\begin{eqnarray}
p(r) =
\f{1}{4}\left(\f{r_0}{r}\right)^{2\alpha}\f{s_0}{r^2}\left[\left(1-2\alpha\right)
s_0 + \f{2}{3}\left(1-\f{2}{3}\alpha\right)\f{s_1}{r}+
...\right],\cr ~\label{eq20}
\end{eqnarray}
where we have expanded all coefficients involving $\alpha$ and
kept only the terms linear in it. The equation of state is
barrotropic, $p(\rho)$.  It is obtained by eliminating $r$ between
Eqs. (\ref{eq17}) and (\ref{eq20}).

\section{Inner solutions}
The formalism developed here is for regions external to the
baryonic matter. Exact interior solutions are involved and are not
easily available. The weak field versions, however, can be
obtained by redefining $M$ of Eqs. (\ref {eq14}) and (\ref{eq16})
as the baryonic mass inside the radius $r$. Thus
\begin{eqnarray}
M(r)\al=\al 4\pi\int_0^r \rho_b r^2 dr, \cr
s_0(r)\al=\al\lambda\left[M(r)/M_\odot\right]^{1/2},\cr
 ~\cr
s_1(r)\al=\al 2GM(r)/c^2.\nonumber
\end{eqnarray}
 The use of $ M(r)$ to calculate $s_1$ in the baryonic
 interior is known to GR and to the
 newtonian gravitation. The proof of its use, to
 infer  $s_0(r)$, is involved.  It is the subject of a
 forthcoming paper by \cite{hhs09}. There we also show that the
rotation curves calculated on this premise are as
good as, if not better than, those obtained by other technics.\\

\section{Concluding remarks}
The proposed formalism is a modified GR paradigm or, equivalently,
a dark matter scenario to understand the non classical behavior of
the rotation curves of spiral galaxies.  We approximate the galaxy
by a spherical distribution of baryonic matter, attribute a dark
perfect gas companion to it , and find its size and distribution
by comparing the rotation curve of our hypothetical model with
those of the actual galaxies. However, as long as the dark
companion displays no physical characteristics other than its
gravitation, one has the option to interpret the scenario as an
alternative theory of gravitation.  For example, one may maintain
that the gravitation outside a baryonic sphere is not what Newton
or Schwarzschild profess, but rather what one infers from the
spacetime metric of Eqs. (\ref {eq6}) - (\ref{eq8}).

Regions exterior to the baryonic matter are not dark matter vacua.
 Therefore, the Ricci scalar does not vanish. Its spatial behavior is that of
$\rho_d(r)$ as can be inferred from the contraction of Eq.
(\ref{eq2}).  There are also excess lensings and excess periastron
precessions caused by the dark matter. These are discussed in
\cite{sob1} and \cite{sob2}.

The formalism is good for spherical distributions of baryonic
matters.  An axiomatic generalization to  non spherical
configurations or to many body systems requires further
deliberations and more accurate observational data for guidance.
One might need further postulates not contemplated so far. The
difficulty lies in the fact that there is no superposition
principle to resort to. One may not add the fields of the dark
companions of two separate baryonic systems; for $s_0$ of Eq.
(\ref {eq14}) is not linear in $M$. As a way out we are planning
to  expand an extended non spherical distribution into its
localized mass-multipole moments and see if it is possible to
assign a dark multipole for each baryonic multipole, more or less
in the same way done for spherical distributions.

\end{document}